\documentstyle[preprint,aps]{revtex}
\begin{document}
\draft
\newcommand{\be}{\begin{equation}}
\newcommand{\ee}{\end{equation}}
\newcommand{\bea}{\begin{eqnarray}}
\newcommand{\eea}{\end{eqnarray}}
\newcommand{\ddl}[1]{\stackrel{\leftarrow}{\partial\over\partial#1}}
\newcommand{\ddr}[1]{\stackrel{\rightarrow}{\partial\over\partial #1}}
\newcommand{\ddd}[2]{{\partial^2\over\partial #1\partial #2}}

\title
{Semiclassical Quantisation Using Diffractive Orbits}

\author{N. D. Whelan}
\address{Centre for Chaos and Turbulence Studies,
Niels Bohr Institute,
Blegdamsvej 17, DK-2100, Copenhagen \O, Denmark}
\date{\today}
\maketitle

\begin{abstract}
Diffraction, in the context of semiclassical mechanics, describes the manner
in which quantum mechanics smooths over discontinuities in the classical
mechanics.  An important example is a billiard with sharp corners; its
semiclassical quantisation
requires the inclusion of diffractive
periodic orbits in addition to classical periodic orbits.
In this paper
we construct the corresponding zeta function and apply it to a scattering
problem which has only diffractive periodic orbits. We find that the resonances
are accurately given by the zeros of the diffractive zeta function.
\end{abstract}


%\narrowtext
%\newpage

Periodic orbit theory \cite{gutz} encapsulates the duality between local,
classical information such as the periods, actions and stabilities of periodic
orbits and global quantum information such as the density of states.  Because
of its reliance on classical mechanics, the theory encounters difficulties
whenever the classical mechanics is singular. Examples of singularities
include three-body collisions as in the Helium atom \cite{tanner};
grazing conditions where some trajectories hit a smooth billiard
surface while very close, parallel trajectories do not \cite{creep};
bouncing from a wedge where
trajectories on one side of the vertex are reflected differently from those on
the other side \cite{vwr,ndw,ps}; and, scattering from a point scatterer
\cite{tbp,br} or a magnetic flux line \cite{br} for which trajectories can not
be continued through the discontinuities. In all of these examples, quantum
mechanics smooths over the discontinuities through diffraction.

In what follows, we study the third example - that of
wedges.  If one is interested in finding the trace of the Green
function $g(E)=\mbox{Tr}G(E)$ (and hence the density of states through
$\rho(E)=-\mbox{Im}g(E)/\pi$) of such a system, one must include the
effect of not just classical periodic orbits but also so-called diffractive
orbits \cite{vwr,ndw,ps}.
These are paths which go directly into at least one vertex.  Such
paths obey classical mechanics everywhere but at the vertex.  There one allows
the path to enter and leave the vertex at any angle \cite{kel} by assigning it
an amplitude obtained by comparison with the exact solution of the
quantum scattering problem \cite{somm}.

This leads to the result that the contribution to the
Green function of a diffractive path from point ${\bf q}_A$ to ${\bf q}_B$
via the vertex
${\bf q}_V$ is approximately \cite{vwr,kel}
\be \label{gdiff}
G_{\mbox{d}}({\bf q}_B,{\bf q}_A,k) \approx G_{\mbox{f}}({\bf q}_B,
{\bf q}_V,k)d(\theta,\theta')G_{\mbox{f}}({\bf q}_V,{\bf q}_A,k).
\ee
Henceforth we assume a billiard system in two dimensions so that
$G_{\mbox{f}}({\bf q}_2,{\bf q}_1,k)=-iH_0^{(+)}(k|{\bf q}_2-{\bf q}_1|)/4$.
The diffraction constant is \cite{vwr,ps}
\be \label{diff}
d(\theta,\theta') = -\frac{4\sin{(\pi/\nu)}}{\nu}
\frac{\sin{(\theta/\nu)}\;\;\;\;\;\;\;\;\;\;\;\sin{(\theta'/\nu)}}
{[\cos{(\pi/\nu)}-\cos{((\theta+\theta')/\nu)}]
[\cos{(\pi/\nu)}-\cos{((\theta-\theta')/\nu)}]}.
\ee

The angles $\theta$ and $\theta'$ are the incoming and outgoing angles
relative to the same face of the wedge, although the choice of face is
arbitrary. The wedge is parameterised by
$\nu=\alpha/\pi$ where $\alpha$ is the opening angle of the wedge.
We have assumed
Dirichlet boundary conditions on all surfaces. Note that $d(\theta,\theta')=0$
when $\alpha=\pi/n$. For these special angles, we can continue any trajectory
through the vertex by flipping the wedge $n$ times to cover the plane
\cite{ps}.
In that event, the contribution to the Green function is not diffractive but
geometric with a phase factor of $(-1)^n$ due to the $n$ specular reflections
all trajectories experience in the wedge.
Eq.~(\ref{gdiff}) is approximate and valid for the points $A$ and $B$ far from
the vertex.  The approximation breaks down when $\theta\pm\theta'=\pi$, a
condition which corresponds to the outgoing angle being directly on the border
between a shadowed and an illuminated region as defined by the incoming angle.
We return to this problem later.

As with geometric orbits, the trace of $G$ is a sum over periodic orbits.
A stationary phase evaluation yields the
contribution of a periodic diffractive orbit labelled $\gamma$ as \cite{ndw,ps}
\be \label{gg}
g_\gamma(k) = -\frac{il_\gamma}{2k}\left\{\prod_{j=1}^{\mu_\gamma}
\frac{d_{\gamma,j}}{\sqrt{8\pi kl_{\gamma,j}}}\right\}\exp{\{i(kL_\gamma+
n_\gamma\pi-3\mu_\gamma\pi/4)\}}.
\ee
The diffractive orbit has $\mu_\gamma$ intersections with a
vertex (diffractions) each with a corresponding diffraction constant
$d_{\gamma,j}$ and $n_\gamma$ reflections off straight hard walls.  The total
length of the orbit $L_\gamma = \sum{l_{\gamma,j}}$ is the sum of the lengths
of the diffractive legs along the orbit and $l_\gamma$ is the length of the
corresponding primitive orbit. If the orbit $\gamma$ is itself primitive
then $L_\gamma=l_\gamma$. If $\gamma$
is the m'th repeat of some shorter orbit $\beta$,then $l_\gamma=l_\beta$,
$L_\gamma=ml_\beta$, $n_\gamma = mp_\beta$ and $\mu_\gamma = m\sigma_\beta$.
We then write
\be
g_\gamma = g_{\beta,m} = -\frac{il_\beta}{2k}t_\beta^m
\ee
where
\be \label{weight}
t_\beta = \left\{\prod_{j=1}^{\sigma_\beta}
\frac{d_{\beta,j}}{\sqrt{8\pi kl_{\beta,j}}}\right\}\exp{\{i(kl_\beta+
p_\beta\pi-3\sigma_\beta\pi/4)\}}.
\ee
This follows because the contributions from the various diffractive legs
in Eq.~(\ref{gg}) are multiplicative so that $g_{\beta,m}$ can be
factorised.  This is not true for geometric (nondiffractive) periodic orbits.

As with geometric orbits \cite{vor}, we can
organise the sum over diffractive orbits as a sum over
the primitive diffractive orbits and a sum over the repetitions
\be
g_{\mbox{d}}(k) = \sum_\beta\sum_{m=1}^\infty g_{\beta,m}
= -\frac{i}{2k}\sum_\beta l_\beta\frac{t_\beta}{1-t_\beta}.
\ee
We cast this as a logarithmic derivative by noting that
$\frac{dt_\beta}{dk} = il_\beta t_\beta-\sigma_\beta t_\beta/2k$ and
recognising that the first term dominates in the semiclassical limit.
It follows that
\be
g_{\mbox{d}}(k) \approx
\frac{1}{2k}\frac{d}{dk}\left\{\ln\prod_\beta(1-t_\beta)\right\}.
\ee
In addition to the diffractive orbits, one must also sum over the geometric
(nondiffractive) periodic orbits so that in the logarithmic derivative we
should also multiply by the contributions from the geometric orbits
\cite{creep}.  In what follows, we assume that all periodic orbits are
diffractive so that the poles of $g(k)$ are the zeros of the zeta function
\cite{zet,ce,aac}
\be \label{zetdef}
\zeta^{-1}(k) = \prod_\beta(1-t_\beta).
\ee
For geometric orbits, this is evaluated using a cycle expansion \cite{ce,aac}.
Here the weights $t_\beta$ are multiplicative so the zeta function is a finite
polynomial conveniently represented as the determinant of a Markov graph
\cite{mark}.

It is instructive to consider a system
which can be quantised solely in terms of periodic diffractive orbits, such as
the geometry of Fig.~1. The classical mechanics
consists of free motion followed by specular reflections off the sides of the
wedges. The two vertices are sources of diffraction. The choice
$\gamma_1=\gamma_2$ has been used to study microwave waveguides
and conduction in mesoscopic devices and is known to have at least one
bound state \cite{wg,ws}.
Unfortunately, in this case $\theta+\theta'=\pi$ for
the periodic orbits labelled $B$ and $B'$ in Fig.~1b; Eq.~(\ref{diff})
then diverges and the diffractive picture breaks down, as mentioned above.
Instead, we consider $\gamma_1>\gamma_2$. This is an open system with
no bound states, only scattering resonances. As these are poles
of $g(k)$ and hence zeros of $\zeta^{-1}(k)$ we can test the
effectiveness of the theory in predicting them.

In what follows, we consider the case $\gamma_2 = 0$ for which the
exact results are simple to obtain. (All the analytical results, however, are
valid for $\gamma_2\ne 0$).  We define four cases:
i) $\gamma_1=\pi/2$; ii) $\gamma_1>\pi/2$; iii) $\gamma_1=\pi/2n$ $(n>2)$; and
iv) all other values of $\gamma_1$.  Cases i) and iii) differ from ii) and iv)
respectively in that $\gamma_1$ corresponds to a special geometric angle and
the lower vertex is not a source of diffraction.
The reflection symmetry of the problem implies that all resonances are either
even or odd. As a result, the zeta
function factorises as $\zeta^{-1}=\zeta_+^{-1}\zeta_-^{-1}$ and we
determine $\zeta_+^{-1}$ and $\zeta_-^{-1}$ separately.

For cases i) and ii) there is only one periodic diffractive orbit which is
labelled A in Fig.~1a.
In the first case there is only one diffraction point while in the second case
there are two. The weight of the periodic orbit in the two cases is
found from Eq.~(\ref{weight}) as
$t_A=d_{AA}\exp\{i(2kL+\pi/4)\}/\sqrt{16\pi kL}$ and
$t_A=d_{AA}d_{AA}'\exp\{i(2kL+\pi/2)\}/8\pi kL$ respectively.  The diffraction
constants $d_{AA}$ and $d_{AA}'$ refer to diffraction from the top and bottom
vertices respectively and can be found from Eq.~(\ref{diff}). The
resonances are determined by $t_A=1$ which yields (to leading order)
\cite{ndw}
\bea
i)\;\; kL & \approx & n\pi - \pi/8 - {i\over 2}
\ln\left(\frac{\sqrt{16\pi(n\pi-\pi/8)}}
{d_{AA}}\right). \nonumber\\
ii)\;\;\;  kL & \approx & n\pi - \pi/4 - {i\over 2}
\ln\left(\frac{8\pi(n\pi-\pi/4)}
{d_{AA}d_{AA}'}\right). \label{oneorb}
\eea
where $n$ is a positive integer. Because the sole periodic orbit is on the
symmetry axis, all of the resonances are predicted to be of even parity
\cite{bent}.  Unlike scattering systems quantised with geometric orbits
\cite{creep,ndw,ce,crrv}, here there are no subleading resonances, a fact also
observed in Refs.\cite{ndw,tbp}.
This is a result of the multiplicativity of the Green function (\ref{gdiff}).

For comparison, the exact resonances are found as follows.
Defining polar coordinates with respect to the lower vertex, we expand
the (unnormalisable) resonance wave function as
\bea
\psi(r,\theta) & = & \sum_{n=1}^
\infty a_n\sin{(\alpha_n\theta)}J_{\alpha_n}(kr) \;\;\;\;\; r\le L \nonumber \\
          & = & \sum_{n=1}^
\infty b_n\sin{(\beta_n\theta)}H_{\beta_n}^{(+)}(kr)
\;\;\;\; r\ge L, \label{psi}
\eea
where $\alpha_n=(2n+1)\pi/2\gamma_1$ and $\beta_n=n\pi/\gamma_1$.  This
satisfies all the boundary conditions and we have restricted the discussion to
even resonances.  Demanding that the wavefunction and its normal
derivative be continuous along the arc $r=L$ gives the quantisation condition
$\mbox{det} M=0$ where
\be \label{mnm}
M_{nm} = (-1)^{n+m}{2 \over \pi}{n \over n^2-(m-1/2)^2}W_{nm}(kL)
\ee
and $W_{nm}(z)$ is the Wronskian of $J_{\alpha_m}(z)$ and
$H_{\beta_n}^{(+)}(z)$. This relatively simple solution is a result of taking
$\gamma_2=0$.

The results for both the exact resonances and their semiclassical
approximations are shown in Fig.~2a for $\gamma_1=\pi/2$ and
$\gamma_1=\pi-0.5$.
The weight $t_A$ is greater by $O(\sqrt{kL})$ in the first case so that the
resonances are not as unstable (i.e. not as far down in the complex
$k$ plane.) Although the agreement is very good, the theory is not very rich
since it is based on just one periodic orbit. Of greater interest
is a situation in which
there are more diffractive orbits.  As $\gamma_1$ is decreased, a
pair of diffractive orbits is born each time $\gamma_1$ passes through a
special angle $\pi/2n$. In particular, for $\pi/4\le\gamma_1<\pi/2$ there are
three fundamental periodic orbits \cite{aac}, as sketched in Fig.~1b,
together with an infinite
hierarchy of longer orbits labelled by their itineraries among the points
$A$, $B$ and $B'$.  For the special case $\gamma_1=\pi/4$, the lower vertex
is not diffractive but rather induces two specular reflections.

The weight of any orbit in which
a letter is repeated can be expressed as the product of shorter weights (eg.
$t_{ABB'B}=t_{AB}t_{B'B}$.)  This means that the system can be written as a
finite transfer matrix and the zeta function is the determinant of the
corresponding Markov graph \cite{predpd}.
Due to the parity symmetry of the problem,
we can consider just the right half of Fig.~1b as the fundamental domain.
Fig.~3 is the Markov graph of the system and shows all the ways of connecting
points $A$ and $B$. For example, the
line marked as 1 denotes starting at point $B$, going to the upper vertex
and diffracting back. Symbols with a bar over them correspond to ending on the
left half of the figure and then reflecting back onto the
fundamental domain.  For example, $\bar{2}$ denotes starting at $B$ diffracting
via the upper vertex to $B'$ and then reflecting back onto $B$. Barred symbols
contribute with a relative positive (negative) sign for the even (odd)
resonances \cite{bent}. Symbol 5 is a so-called boundary orbit which lies on
the border of the
fundamental domain and contributes only to the even resonances. Each
symbol carries its own weight and these combine to give the weights of
the periodic orbits.
To find an expression for the zeta function, we enumerate
all non-intersecting closed loops and non-intersecting products of closed loops
to obtain \cite{predpd}
\be \label{zeta}
\zeta^{-1}_{\pm} = 1-t_1\mp t_{\bar{2}}-t_3(t_4\pm t_{\bar{4}}) -
\frac{1\pm 1}{2}(t_5-t_5t_1\mp t_5t_{\bar{2}}).
\ee
Since by symmetry $t_4=t_{\bar{4}}$, we have
\bea
\zeta^{-1}_+ & = &
1-t_1-t_{\bar{2}}-t_5-2t_3t_4+t_5(t_1+t_{\bar{2}}) \nonumber \\
\zeta^{-1}_- & = & 1-(t_1-t_{\bar{2}}). \label{zetapm}
\eea

We must still define the weights which appear in these formulas. Each symbol
is composed of two segments leading from the corresponding nodes to the
vertex connected by one diffraction at the vertex. We make use of this
fact by separately defining quantities which contain the information
about the segments and about the diffractions. The segment information is
contained in $u_A$ and $u_B$ which are given by
$u_A^2 = d_{AA}'\exp\{i(2kL-3\pi/4)\}/8\pi kL$ for
$\gamma_1>\pi/4$ or $u_A^2 = \exp\{i2kL\}/\sqrt{16\pi kL}$ for
$\gamma_1=\pi/4$, and $u_B^2=\exp\{i(2kH+\pi)\}/\sqrt{16\pi kH}$.  These are
simply the square roots of the weights of the fundamental periodic orbits shown
in Fig.~1b but without the phase and diffraction constant from the vertex.
That information we quantify by defining the phase factor
$s=\exp\{-i3\pi/4\}$ and four diffraction constants. There is a constant to
diffract from $A$ back to itself which for $\gamma_2=0$ is $d_{AA}=2$.
With similar notation we find $d_{AB}=2\csc(\gamma_1/2+\pi/4)$ and
$d_{BB}=d_{BB'}=1+\csc\gamma_1$. (Note that by symmetry
$d_{ij}=d_{ji}$.) The equality of $d_{BB}$ and $d_{BB'}$ is an artifact of
choosing $\gamma_2=0$ and is not a general result. We then have
$t_1=sd_{BB}u_B^2$,
$t_{\bar{2}}=sd_{BB'}u_B^2$, $t_3=t_4=t_{\bar{4}}=sd_{AB}u_Au_B$ and
$t_5=sd_{AA}u_A^2$.

We stress that the only difference between cases iii) and iv) is the form of
$u_A$ and that the functional form (\ref{zeta}) applies to both.  However, as
we will see, the numerical results are quite different. Due to the equality
of $d_{BB}$ and $d_{BB'}$, the weights $t_1$ and $t_{\bar{2}}$ are equal so
there are no odd resonances.  This can also be seen from the fact that
$J_\nu(z)$ and $H_\nu^{(+)}(z)$ are independent functions so there can be no
odd resonances which satisfy all the boundary conditions and match smoothly
at $r=L$.

We show the results for the even resonances for the cases $\gamma_1=\pi/3$ and
$\gamma_1=\pi/4$ in Figs.~2b and 2c respectively.
In the first case there is a diffraction at the
lower vertex so that $u_A/u_B=O(1/\sqrt{kL})$.  Therefore the dominant
behaviour for $\zeta_+^{-1}$ is dictated by $t_1$ and $t_{\bar{2}}$ and the
other terms represent small corrections.  This is apparent in the figure where
the gross behaviour is similar to Fig.~2a but with a decreasing, oscillatory
correction coming from $u_A$.  In the second case $u_A$ is of the same order
as $u_B$ so that $t_1$, $t_{\bar{2}}$ and $t_5$ are all of similar magnitude
Then the resonances have no obvious pattern but are
scattered around the complex $k$ plane
because of strong interferences among the various terms in  $\zeta_+^{-1}$.

In both cases the semiclassical approximation is very accurate.
In addition it
gives us a better qualitative understanding of the spectrum than the exact
numerical calculation based on Eq.~(\ref{psi}).  Although we can not
yet study the case $\gamma_1=\gamma_2$ in detail, we can begin to understand
why none of its bound states are odd \cite{ws}.
Recall that bound states correspond to real zeros of the zeta functions.
However, the odd
zeta function receives no contribution from the boundary orbit and also
suffers from strong cancellation due to the relative negative sign between
$t_1$ and $t_2$, so only quite far from the real $k$ axis
are the magnitudes of the weights sufficiently large to allow for a zero.
It is reasonable to suppose that this general condition
continues to hold as $\gamma_1\rightarrow\gamma_2$. However, a complete
semiclassical analysis of this case requires an understanding of the
behaviour of the Green function on the border between shadowed and
illuminated regions.

Finally, we mention that the restriction to $\gamma_1\geq\pi/4$ is not
crucial.  For smaller values of the angle, more diffractive orbits appear but
the formalism above still applies and can be used to find the corresponding
zeta function.

I thank Stephen Creagh,  Predrag Cvitanovi\'{c}, Peter Dimon,
Julie Lefebvre, Gregor Tanner, G\'{a}bor Vattay and Andreas Wirzba for useful
discussions.
This work was supported by the EU Human Capital and Mobility Programme.

\begin{figure}
\caption{a) A configuration with only one diffractive periodic
orbit. b) A configuration with an infinity of diffractive periodic
orbits labelled by their itineraries among $A$, $B$ and $B'$.}
\end{figure}

\begin{figure}
\caption{The resonances in the complex $k$ plane. a) Results with only one
diffractive periodic orbit.  The upper set of points
corresponds to $\gamma_1=\pi/2$ and the lower set corresponds to
$\gamma_1=\pi-0.5$.  For each set, the exact resonances are denoted with
crosses and the semiclassical approximations with diamonds.
With the same convention, b) shows the results for
$\gamma_1=\pi/3$ and c) shows the results for $\gamma_1=\pi/4$.}
\end{figure}

\begin{figure}
\caption{The Markov graph used to compute the desymmetrised zeta functions.}
\end{figure}

\end{document}